\documentclass[11pt,a4paper]{article}

\usepackage[utf8]{inputenc}
\usepackage[T1]{fontenc}

\usepackage[margin=1in, top=0.75in]{geometry}

\usepackage{amsmath,amssymb,amsfonts}

\usepackage{graphicx}

\usepackage[colorlinks=true,linkcolor=blue,citecolor=blue,urlcolor=blue]{hyperref}

\usepackage{epigraph}
\usepackage[round,authoryear]{natbib}
\usepackage{setspace}
\usepackage{fancyhdr}
\fancypagestyle{firstpage}{
  \fancyhf{}

  \fancyfoot[C]{\footnotesize\textit{Preprint (not peer-reviewed) $\cdot$ February 3, 2026 $\cdot$ \textcopyright\ The authors/funders $\cdot$ CC-BY-NC-ND 4.0}}
}
\fancypagestyle{plain}{
  \fancyhf{}
  
  \fancyfoot[C]{\thepage}
}

\usepackage{titlesec}
\titleformat{\section}{\large\bfseries}{\thesection}{1em}{}
\titleformat{\subsection}{\normalsize\bfseries}{\thesubsection}{1em}{}

\usepackage{authblk}

\begin{document}

\title{\textbf{Reshaping Perception Through Technology: From Ancient Script to Large Language Models}}

\author[1,2]{Parham Pourdavood}
\author[1,2]{Michael Jacob}

\affil[1]{Department of Psychiatry and Weill Institute for Neurosciences, University of California, San Francisco, 505 Parnassus Ave, San Francisco, CA 94143, United States}
\affil[2]{Mental Health Service, San Francisco VA Medical Center, 4150 Clement St, San Francisco, CA 94121, United States}

\date{}

\maketitle
\thispagestyle{firstpage}

\noindent\textbf{Correspondence:}\\
Michael.Jacob@ucsf.edu\\
Parham.Pourdavood@ucsf.edu

\vspace{1em}

\begin{abstract}
As large language models reshape how we create and access information, questions arise about how to frame their role in human creative and cognitive life. We argue that AI is best understood not as artificial intelligence but as a new medium---one that, like writing before it, reshapes perception and enables novel forms of creativity. Drawing on Marshall McLuhan's insight that ``the medium is the massage,'' we trace a lineage of technologies---from DNA and the nervous system to symbols, writing, and now LLMs---that mold cognition through a shared logic of flexible unfolding and co-creation. We observe that as technologies become more externalized and decoupled from physiology, they introduce both greater creative potential and greater risk of inauthenticity and manipulation. This tension is acute with LLMs, but not unprecedented: ancient responses to writing reveal a recurring human tendency to project intelligence onto powerful new media. Rather than viewing AI as a competitor, we propose framing it as a medium that foregrounds artistic skills: aesthetic judgment, curation, and the articulation of vision. We discuss implications for education, creative practice, and how society might adapt to this new medium as it did to writing.
\end{abstract}

\section{Introduction: The Felt Dimension of Writing}

\epigraph{``The painter's products stand before us as though they were alive. But if you question them, they maintain a most majestic silence. It is the same with written words. They seem to talk to you as though they were intelligent, but if you ask them anything about what they say from a desire to be instructed they go on telling just the same thing forever.''}{ -- Plato, \textit{Phaedrus}}

Writing is typically regarded as an output or expression of human consciousness -- thoughts made visible, a mirror of inner experience. Given this perspective, revisiting writing and its invention may seem like putting the cart before the horse, in a study of consciousness. However, Plato's observation about writing's peculiar pseudo-intelligence suggests a more complex and nuanced relationship to this medium. We readily acknowledge that the arts and humanities can ``elevate'' or transform our consciousness, providing new perspectives and experiences. But perhaps the transformative power of writing, the arts, and technology more broadly, can tell us about the nature of consciousness -- precisely because these mediums can so profoundly alter it.

The ancient Greek philosophers had mixed feelings about the revolutionary technology of their time, writing. Among their major concerns were writing's lack of living context and dialectic, as well as a degenerating effect on the memory of its users (``forgetfulness of the soul'', as Socrates echoed through Plato). Paradoxically, the realization of this lack of living context and dynamism coexisted with projections of intelligence onto written texts, with a tinge of artificiality and magic \citep{hackforth1972plato}. Similar to Plato's account of written words that ``talk to you as though they were intelligent'', American philosopher David Abram, in his book \textit{The Spell of the Sensuous}, provides historical accounts of how indigenous tribes from different continents independently described the written texts Europeans engaged with as ``talking leaves''. These magical origins are etymologically preserved, Abram notes in our English word \textit{spell} -- that we now recognize as an arrangement of letters in correct order -- itself borrowed its significance from the concept of magical spells \citep{abram2012spell}.

So, why did a medium such as writing -- which has become a trivial part of our day-to-day life -- generate such an eerie, if not unsettling, enchantment in its early users? Is there a general principle for how externalized creations reshape the conscious experience of their creators? In an apparent paradox, we argue that understanding consciousness, that most private of inner experiences, requires turning our attention to the cultural artifacts it produces.

We now realize that the feeling of magic induced by writing was not a property of writing but how consciousness was developing in real time as a result of encountering a new informational medium. The magic and intelligence that early adopters of writing experienced were the felt effects of perception, differentiating and venturing into new territories.

This is perhaps the most evident it has been since the invention of writing, in how we have framed the revolutionary technology of our time as ``artificial intelligence''. The attribution of intelligence to these systems is reminiscent of attributions of intelligence to the invention of writing during antiquity. A lesson we can learn from that era is the need to move beyond projecting our aliveness and intelligence onto our creations and instead focus on how these externalized technologies -- such as symbols, writing, and AI -- can become harmonious and naturalizable extensions of our cognitive milieu. And further, our entanglement with them may differentiate our consciousness and creative capacity, even as we become increasingly dependent on them.

While at first blush this loop may appear unique to human technology, closer examination suggests it is in fact resonant with the fundamental principles of biology, our brains, and perhaps, consciousness itself. The evolution of life can be seen as a process of flexible unfolding into an ever-expanding world. From our DNA, to the cells of an embryo, to our nervous systems, and even into new tools such as writing and AI -- life encounters new architectures. Through these architectures, life flexibly and vulnerably reshapes its unfolding process to accommodate and find value in them -- to eventually ``fold them in'' as a way of learning and consolidation. Life thrives on playful experimentation and openness to change, but it needs the successful learned records of these playful and flexible dynamics to pass them on and draw from to further its ongoing creativity. Moreover, this entanglement with and appreciation of a new medium enables novel ways of creativity and expression that organisms need to adopt and adapt to. And as we will show, this novel work of unfolding and reshaping as a result of encountering unique mediums is accompanied by a novel sense of perception and felt experience. Focusing on how consciousness develops and expands as it faces novel situations offers a clue as to what it is. As conscious experience is increasingly defined by processes and not things, how it develops may help us understand what it ultimately is.

\section{Life Unfolds, So Does Your Consciousness}

It's hard to imagine a biology textbook without an early discussion of circuits and blueprints. Such language, and the corresponding flowcharts, give the impression of an engineered logic. However, lost in the complexity is the fact that structures of biology are not designed, nor constructed -- they organically develop and evolve; they unfold.

Consider embryology: life begins with a single cell (the fertilized egg) while containing the potential to create the full form and dynamics of an adult organism (think of it as an extremely folded origami). What follows in the developmental process is a remarkable process of successive unfoldings and differentiation, while at each step remaining a working whole.

It is natural to ask then: how did things get folded in the first place? And what would it even mean to suggest that life is folded? Here, we consider unfolding not only to capture the process of differentiation, but to emphasize the physicality of life's interactions, and the necessity of finding means to record useful ones. In a certain sense, life needed to establish a system like writing before evolution could proceed. Something that could record information about life's ingredients and recipes, pass these records forward from generation to generation, and establish new variations on those recipes.

Of course, that something is DNA. We take its existence for granted, but DNA, too, had to evolve in the earliest stages of life's history on Earth. DNA is literally folded up, long strands wrapped around larger molecules called histones. And although we speak of information being ``encoded'' in DNA, this information is never actually transmitted, like digital code through hardware, but depends on the geometry of unfolding the layers of this helical structure in the presence of other molecules. In other words, when a potentially useful variation in the genetic code is ``discovered'' through evolution, this discovery is folded into the lineage of a species so that it can be later unfolded during development. Life could be seen as one long evolutionary dance composed of the work of folding and unfolding -- this brings new usable and playable motifs in contact. Moreover, DNA required a restructuring of how organisms evolve, develop, collaborate, and communicate.

Finally, a subtle yet important point is that folding and unfolding are by nature geometrical and physical processes, not computational. Folding requires geometry, and the only source of geometry is physical forms. By emphasizing the physical interactions required for folding and unfolding, we do not mean to imply that consciousness is ``stuff.'' Instead, we bring attention to the importance and effect of geometry on the dynamics of life, and the unique contexts in which new geometric relations are brought into contact with during folding and unfolding.

\section{Folding and Unfolding in the Brain}

A particularly remarkable feature of biological unfolding is its flexibility. Despite terminology like ``developmental program,'' organisms can adapt to continuously changing circumstances, sometimes dramatically. This is evident in embryology, where the organism ``adapts to the body it finds itself in'', regardless of how unprecedented the context might be. In a striking example, grafting a third eye into a frog embryo results in the frog developing binocular vision maps in their brain -- a phylogenetically unprecedented function in frogs. The frog reshaped its unfolding dynamics to accommodate the new eye and, as a result, found an appreciation for its use.

As we see it, biological organisms have been habituated by evolution to realize any new pattern that comes their way as a potential, meaningful mini ``origami'' that can be unfolded as part of their ongoing dynamics. The brain exemplifies this capacity in spades; neuroplasticity is a defining feature of the mammalian nervous system. In fact, it is the capacity of neuronal connections, synapses, to physically rewire on the basis of new information that enables the frog's third eye to be meaningfully incorporated into the frog's perceptual apparatus.

We agree with philosopher Thomas Nagel's famous essay conclusion that there must be something it is like to be a bat, or a frog, for that matter \citep{nagel1974bat}. And while it is not possible for us to conceive of how the addition of a third eye might have altered the experience of the frog, it seems likely that it did. This is because we know from decades of work in neuroscience that changes in the wiring of the brain are meaningfully correlated with changes in the structure of experience. Neurobiology should no doubt give us important clues about how the folding and unfolding of this particular information medium is linked to the experience of consciousness. But to understand the information medium of the brain, we need to examine not only the structure of its synapses, but its dynamical, electrical activity.

In a dramatic reversal of the standard logic of electrical engineering, brain activity never ``turns-off.'' Across states as varied as sleep, daydreaming, or philosophical contemplation, the brain sustains both constant energy consumption and richly abundant, patterned electrical activity. Incessant neuro-electrical dynamics mirror fluid dynamics, with signals cascading through increasingly specialized networks. What is the meaning of this constant activity?

For decades, this so-called ``spontaneous activity'' was dismissed as mere idling ``noise.'' But developmental neuroscientists recognized that a sufficient amount of internally generated activity was crucial for embryological development. This insight provides an important clue toward the significance of this tireless activity: constant dynamics affords generativity and serendipitous synergies that serve as the prima materia for unfolding of and fixing perception and action. Ongoing brain activity is a motivated spontaneity.

In the brain, spontaneous dynamics affords dynamical shaping of the electrical activity itself by introducing constant novelty -- and that enables the very unfolding process to recognize the effects and consequences of novel, spontaneous activities. Since the brain is constantly active and generating spontaneous variations, developing its dynamics is more like carving and guiding its activity to find usable variations, rather than constructing them anew from the ground up. For this reason, brain processes have been described as ``auto-sculpture,'' with perceptual input ``sculpting'' away at the ongoing activity, which is then further consolidated through ongoing action and engagement in the world. The history of these useful consolidations then become stable and available for future use in synaptic strengths \citep{uddin2020noise}.

Given this understanding, brain dynamics more closely mirror the embryologic logic of differentiation and flexible unfolding described above than the operation of an electrical circuit. Classical research in perceptual decision-making reveals that neural activity accumulates over time as organisms prepare to act. This accumulation is driven by motivation but refined through inhibitory signals from neurons associated with competing or irrelevant decision options. Critically, the spontaneous variability inherent in this process affords the potential for the exploration of novel patterns, and through neuroplasticity, the potential to write and rewrite the synaptic structures that channel neuro-electrical activity in the first place. Seen in this light, our neural network is a ``written artifact'', representing the history of useful sculptings of an endless ocean of spontaneous activity -- the manifestation of evolutionary, developmental, and experiential unfolding. This written artifact is similar to what German philosopher Friedrich Schelling called architecture as ``frozen music'' -- in our example, the music is the ongoing spontaneous activity itself and the synaptic strengths the resultant architecture \citep{schelling1845philosophy}.

These principles, such as becoming a written artifact of useful spontaneous behavior and plasticity, are not limited to physical implants during development either, as was the case with the frog example. Consider the famous Rubber Hand Illusion experiment, where a participant's brain is deceived by covering one of their hands while placing a fake rubber hand in their visual field. After, for example, stroking a ruler a few times on both hands in the same direction, the user starts reporting sensations of false ownership, anticipation of pain when threatened to get hit, and even subtle tingling in the fake hand as if it were real. This extension of experience onto an externalized medium (like the ``magic'' and ``intelligence'' of written words) is a demonstration of flexible neuroplasticity and fitting to spontaneous novelty: we do not just react to stimuli and receive feedback errors (like machine learning systems do) but the novel environments we interact with make us ``fall out'' of our ordinary shape to accommodate them and appreciate their effects. Importantly, this reshaping is not merely an abstract, informational correspondence, but it requires meaningful entanglement with the organic processes of life \citep{botvinick1998rubber}.

From DNA to brains, we need to reconsider the meaning of information in life and consciousness. Through life, consciousness expresses creative unfoldings into new places, accruing new insights about the world. These insights leave a mark, an imprint on us and we can see this, quite literally in the structure of our brains and nervous systems that mediate the leading edge of this exploration.

\section{The Medium Shapes the Message}

This discussion highlights our emphasis on the word \textit{medium}: the defining mediation between subject and object. A medium's spatiotemporal features are key in this mediation process. Canadian philosopher Marshall McLuhan famously coined the phrase, ``The medium is the message'', observing that the unique properties associated with a novel medium profoundly shape our conscious experience -- they literally become stretched ``extensions'' of our perceptual process, as he put it. Consequently, the real message is not a code in a channel of information, but rather how a medium's unique characteristics directly shape our experience from the ground up. For this reason, McLuhan later adopted the title ``the medium is the massage'', to imply how the unique features of mediums ``massage'' our conscious experience \citep{mcluhan1967medium}.

The concept of reshaping as a result of unfolding physical forms hints at the work of early American pragmatists. John Dewey, a key figure in this movement, emphasized ``learning by doing'', that learned value is the result of physical engagement with the world; mentation in effect is the shape of doing, not abstract manipulation and transmission of knowledge \citep{dewey2005art}.

In a similar vein, Charles Sanders Peirce, who coined the term pragmatism, emphasized that our best shot at understanding the meaning and significance of any object is recognizing and mediating (what we have framed as contextual unfolding) its relevant physical features and consequences. For instance, remember how earlier in our discussion, a grafted third eye for the developing frog meant depth perception because that's how the frog experienced the practical effects of the added eye (e.g., the novel geometry it introduced) as part of its ongoing dynamical embodiment.

It should be noted that pragmatism does not suggest that perception is necessarily correlated with usefulness,\footnote{This common misunderstanding forced Peirce to change the name of his theory to pragmaticism -- a name ``ugly enough to be safe from kidnappers'' (Collected Papers (CP) 5.414).} but that any sort of experienced significance (useful or not) emerges through recognizing a medium's physical features relative to the contextual milieu of oneself (consider how the effects of a chemical substance can be experienced as lethal poison to one organism and delicious food to another one). In general, meaning in organisms is communicated and contextually recognized through the geometric features of a diverse set of internal and external mediums (e.g., the 3D shape of hormones, amino acids, RNA, pheromones, viral proteins, etc.).

Examining this issue a bit more carefully, we find that the evolution of organisms in general can be thought of as serendipitous encounters with novel patterns and informational mediums that became sedimented as part of the intrinsic unfolding of their effects.\footnote{Biologists refer to this process as ``evo-devo,'' reflecting how the plasticity of developmental processes can provide a source for variation in evolution.} The genetic system, the nervous system, the immune system, etc, are all examples of informational mediums that the organism learned to adapt to, use their organizational capacities, become creative upon, and get entangled within a process of co-evolution and co-creation. By creation, we mean the creative capacity of all life: the potential to realize the relevance of new patterns and rewrite them into their own life fabric. And as the organism adopts these novel patterns, its interpretation -- and possibly recreation -- of them changes accordingly.

This general logic of innovation by co-option is part of a general tendency of organisms to remain evolvable, which is defined as not only adapting and surviving but complexifying and flourishing by constantly navigating their useful variations. Evolutionary theorists refer to this process of co-opting as exaptation -- one a trait that evolved for one purpose, becomes transformed into another. Recombining and repurposing of functions is nature's most efficient and flexible way of remaining evolvable. Constant innovation is necessary in the face of constant degradation and change imposed by nature; metaphorically, the only way to remain in one place is to keep moving.\footnote{This is famously known as the ``Red Queen's Hypothesis''.} This is the reason why life needs to remain open to the effects of architectures it comes across, in spite of the uncertainty and risk they introduce.

And while this capacity began with life, and has been extended through the nervous system, it is now further extended into an increasingly wide range of cultural artifacts. Therefore, our consciousness too is not only entangled with our biological and cultural heritage, but itself embodies an evolutionary logic of innovation and unfolding.

The takeaway from this biological argument is that evolution has ingrained in organisms a habit of being reshaped as a result of flexibly recognizing the effects of novel, spontaneous patterns they come across. Living organisms demand a habit and inclination for innovating upon novel artifacts that they get entangled with in relation to their becoming. As Antoine de Saint-Exup\'{e}ry noted, ``If you want to build a ship, don't drum up people to collect wood and don't assign them tasks and work, but rather teach them to long for the endless immensity of the sea.'' Purposeful ``longing'' enables organisms to afford vulnerable openness to the creative value of novel mediums they unexpectedly become entangled with along the way, while still remaining grounded to their general goals. But, this is still a vulnerable habit since unfolding of unknown possibilities in a dynamic environment risks unintended consequences.

Consider an inventor obsessed with flight in the early 1900s -- there are many ways to slice the problem of getting airborne. You could study bird wings, experiment with lighter-than-air gases or build gliders. Each approach creates its own set of mechanical tensions -- lift versus weight, power versus control, stability versus maneuverability. Longing keeps pulling the inventor through the unfolding explorations that are necessary to resolve each of these unique tensions, incorporating new insights along the way. For Wilbur Wright, it was simply the ``desire to fly'' and soar ``in the infinite highway of the air.'' Most inventions reflect this intersection of a medium in a new context and the vital significance of longing. Feeling a longing is the easiest way to enforce it. Ongoing attempts to resolve the tensions created by novel patterns are motivated and felt by us organisms to always contextually reorient us towards our overarching goals. So, consciousness grows with re-cognition\footnote{The dash in recognition is meant to emphasize the notion of ``seeing in a new way''; finding value in spontaneous mediums is a transformation of cognition.} of value and felt tension in these patterns, and as encounters with novel information become more complex, so does the organism's competence to see the world in a new way -- and so does consciousness. We may not know exactly where this exploration will take us, what the experience of the new medium will yield, but we are carried along by the vital longing of life for creative appreciation.

\section{Creative Consciousness}

To make our hypothesis as explicit as possible, conscious experience is always in an artistic process of unfolding. This process of recognizing novel and valued meanings is at the root of any creative act. As John Dewey said in his book \textit{Art as Experience}:

\begin{quote}
``The work is artistic [\ldots]. As the painter places pigment upon the canvas, or imagines it placed there, his ideas and feeling are also ordered. As the writer composes in his medium of words what he wants to say, his idea takes on for himself perceptible form''.
\end{quote}

\noindent In this way, the human artistic process can be thought of as a generative process of constantly reshaping as a result of interpreting our own creations. As the artist starts to create new patterns, they introduce irritations and tensions within themselves, and the feeling of those tensions guides further aesthetic shaping. This unfolding of one's creations develops through a conscious experience of entangled co-creation, motivated by risk and tension, until satisfaction is achieved.

A play of expressive exploration and realized satisfaction mirrors the sculpting dynamics of the nervous system that unfolds through motivated spontaneity. The process itself is messy, risky, and vulnerable. The root of the word \textit{experience} in English is itself noteworthy, which has its roots in the Latin word \textit{ex-periri}, which means ``to try'' and ``to risk''. This linguistic hint might suggest that going beyond the familiar to recognize something new inherently feels like a risk.

Risk is at the heart of any creative work, although subtly: disturbing, offsetting, and recombining what you already know is the only way for novelty to emerge. Venturing into the unknown and being creative is inherently a self-doubting and vulnerable act, but the reward for this irritation and ambivalence is aesthetic novelty and growth. So, it is not just that consciousness creatively produces art, but how consciousness develops is itself a creative and artistic process of vulnerably accepting novel patterns and being reshaped as a result.

For an example of a risky, artistic adventure leading to differentiation of perception, consider musical temperament. In traditional pure tuning systems, the distances between musical notes are based on ``perfect'' whole ratios (two to one, three to two, etc). In this system, even though the intervals sound perfectly consonant and harmonious in a specific key, moving across keys (which is known as modulation in music theory) becomes a challenge. Musical temperament solved this issue by messing up the ``perfectness'' of intervals ever so slightly as a compromise so that one can have reasonable intervals in all keys and hence can move between keys seamlessly.

A surprising result of this tempering was that different keys acquired unique affective and emotional qualities for listeners (e.g., C Major felt innocent, D Minor felt melancholy, F Major felt compassionate, F Minor felt like a funeral lament, etc.). Johann Sebastian Bach's use of Well Temperament in his monumental work, \textit{The Well-Tempered Clavier}, was motivated by composing pieces in all twenty four distinct keys, while showcasing their different emotional qualities. Creating different temperament systems to get unique qualities itself became an artistic expression for musicians, playing with the boundaries of our conscious perception \citep{schubart1806ideen}.

This novel medium also meant that the music practice and community as a whole had to adapt and learn new ways of creativity and expression: instrument-makers redesigning their craft, teachers learning how to teach music with well-temperament, composers learning how to compose in all keys, and so on. This shows how quickly the effects of a novel medium can reverberate into society as a whole and change humanity's collective perception.

\section{Shaped by a Sea of Symbols}

Symbols are an example of an externalized medium that humans have become entangled with and that has dramatically reshaped our consciousness. In fact, the evolution of homo sapiens has been framed as a co-evolution between humans and symbols. Neuroanthropologist Terrence Deacon, in his book titled \textit{The Symbolic Species: The Co-evolution of Language and the Brain}, explains how the brain had to restructure itself in order to host symbolic thinking. This is analogous to how dams restructured how beavers function and go about their lives, what in evolutionary theory is referred to as niche-construction. Symbols required a reconstruction of how we experience life as individuals and as a society, because we could now reference experience beyond the here-and-now. Symbols depend on social and conventional relationships and can only be understood systematically with each other. This opened up an entirely new mode of communication that life had never experienced before, changing our physiology in order to generate and interpret them, as well as our social and cultural systems in order to accommodate them and pass them on \citep{deacon1998symbolic}.

Symbols opened up novel pathways for creativity and expression by allowing us to conceive of possibilities that are not limited to the immediate: a primate started imagining and talking about the future, the past, the impossible, what-ifs, lies, paradoxes, etc. Symbols provided humans a powerful medium for problem solving, by introducing advanced forms of communication and planning, as well as creative and artistic expression (such as epic poetry, music, dance, storytelling, and rituals).

However, creative imagination is a double-edged sword: symbols can also induce stress and vulnerability -- as we saw with other novel mediums -- since they allow us to think about scenarios or generate narratives that have no basis in reality, such as delusional thinking, irrational fears, propaganda, and manipulation. Adopting symbols was perhaps a risky and extremely unlikely effort, which is evident in the fact that it has happened only to one species, at one point in time during the evolutionary history of life on earth.

What do symbols then mean for conscious experience? Even though the framing of human language is usually focused on its computational features and abilities (it being merely a universal grammar system), how it has shaped human conscious experience remains less known. A scientific attempt that gets close to this endeavor can be found in research on linguistic relativity, namely the Sapir-Whorf hypothesis, proposed by Edward Sapir and further developed by Benjamin Whorf. Put simply, the Sapir-Whorf hypothesis states that the structure of different languages affects the perception and categorization of the experience of their speakers.

There have been numerous studies exploring this hypothesis. A unique line of research has been exploring how languages with different vocabularies for different colors affect boundaries of conscious perception. Classic cross-linguistic research by Kay and Kempton comparing English and Tarahumara speakers has shown that in languages where there are no distinct words for green and blue (sometimes known as \textit{grue} in linguistics), speakers draw less sharp boundaries between what we more decisively perceive as blue and green; the acuity around that color spectrum seems to be more fluid and less crystallized. Even though the wavelengths are registered similarly on their retinas, the resolution of conscious perception develops and differentiates uniquely for speakers with different color categorizations in their language \citep{kay1984sapir}. This hints at a flexible unfolding process being at play and the medium shaping the message. Therefore, it's not just that language is an innate universal grammar being installed alongside our thoughts (as Noam Chomsky suggested), but what thinking itself feels like is shaped by the structures embedded in the unique symbolic frames one adapts to and is engaged with throughout life.

\section{The Artistic Canvas of Writing}

As we mentioned at the beginning of the essay, writing had a profound emotional impact on its early users. Socrates deemed written words to be risky ``external marks'' that ``implant forgetfulness in the soul'' and ``talk to you as though they were intelligent''; indigenous communities associated a feeling of ``magic'' with written words and described them as ``talking leaves''. Following the now familiar, the medium shapes the message argument, the novel feelings and emotions early adopters of writing felt could be framed as conscious perception being vulnerably reshaped as a result of unfolding the novel patterns of a potentially useful medium.

It is obvious that writing is not intelligent and does not really talk to us, but it has become a useful instrument for creativity and expression, and a natural extension of our cognition. And by now, the whole society has restructured itself to harmoniously work and collaborate through this externalized medium. Even though the intuition of Socrates about the deterioration of memory as a result of adopting writing was correct to some extent, society has since come up with new ways of creative and cognitive work (writing books, composing music, creating poetry, writing a screenplay, etc) that are still challenging, but require a different way of seeing creativity than Socrates could have conceived of.

However, the double-edged aspect of creative expression and imagination that we discussed around symbols are inherited in writing. With writing one can spread lasting misinformation, organize cults, rewrite history, or assert control and dominance. However, it can also mobilize for justice, enable collaboration across time and space, and cultivate imagination and empathy through novel arts such as literature and poetry. Regardless of whether the benefits of writing outweigh its harms, its means of organizing our ever-accumulating knowledge seemed inevitable, and gave us a point of no return. Coming up with novel ways to harness its usable effects seemed to be the most pragmatic approach and remedy to offset its dangers.

Writing is a flexible, externalized medium, so it lifted the more labor-heavy and immediate aspect of oral communication and discourse, and allowed people to be creative and express themselves in a low-stakes, efficient medium. It also enabled individuals to pass down their sense of aesthetics, thoughts, and expressions in a way that transcends their physical, mortal body. Writing allowed the society as a whole to experiment and play more efficiently with their collective thoughts and ideas, and hence be more artistic.

During writing one can recombine sentences, use metaphors and drawings, create vivid images and build worlds in complex narratives that readers can get lost in and be emotionally attached to and be moved in a way that would be almost impossible if not for reading. The defining sign of intelligence and scholarship for millennia -- having a remarkable memory -- now seems insignificant compared to the creative potential and value that the medium of writing has opened up.

Creating and consuming bolder, more imaginative works of expression and art inevitably develops and changes our consciousness. Although we can't always put our finger on the unique perceptual territories we travel in as we read or write, we are perhaps constantly affected by tiny ``Sapir-Whorf'' effects as we read a new story or enjoy a piece of poetry. Each unique piece of writing has a specific structure, and our conscious experience's ongoing unfolding process inevitably reshapes as a result of encountering novel forms. The structures that we can now embed in the medium of writing are more flexible and complex than the symbolic medium, so our consciousness grows more intricately as a result of mediating them.

\section{AI: Art in a New Key}

The fact that to its early adopters, writing felt like a magical, intelligent being that would talk to them makes one wonder whether that is why our society is currently feeling the same way regarding the magical medium of our time: artificial intelligence. It is, however, becoming increasingly evident that the current AI models, called large language models (LLMs), are not intelligent, but are providing us a novel, even more powerful medium of artistic expression and creativity than writing. And by doing so, they will shape the conscious experience of individuals and society due to the unique ways and works of creativity and expression they introduce.

Let's analyze whether the argument for AI being a creative canvas for artistic expression and collaboration holds up. A clear observation is that LLMs are lifting mechanical barriers to creativity even more dramatically than writing did. The amount of generativity that one can get with a single click or prompt is unprecedented. We can instantaneously ask questions about the whole of human knowledge and get responses in no time, without turning a single page of a book. We can combine concepts and translate across concepts without having developed expertise in either of them. A novice storyteller can create impressive characters from cultures they haven't had much direct exposure to yet. A musician can blend their genre of expertise with another genre that they don't have much experience in. A physicist can learn about finance by translating financial markets into the language of thermodynamics, and so on. By lifting these mechanical labors, LLMs are enabling us to focus on what we find tasteful and beautiful. They are novel instruments allowing us to do a new kind of art.

The evidence for this new kind of art is emerging in the field of technology itself where software engineering job interviews are increasingly shifting towards asking questions related to design and aesthetics, as opposed to focusing heavily on writing code, which LLMs are becoming increasingly better at. Trendy practices and software under the term ``vibe coding'' are allowing people without any programming background to iteratively communicate their intentions in natural language to LLMs until they get a result they are satisfied with. Since the generation of artificial novelty and acquiring knowledge have become trivial, the talent that is increasingly becoming valuable is aesthetic judgment, curation/recombination of forms and ideas, and clearly expressing your vision. This kind of talent, however, is usually regarded as the most recognized position in artistic projects: film directors and music conductors don't mechanically produce something; these individuals have worked to develop a great sense of interpretation, aesthetic judgment, and the ability to express their taste.

Similar to the dangers of symbols and writing, the negative dimension of the creativity and generativity that LLMs introduce is going to exacerbate issues such as misinformation, delusional thinking, addiction, and propaganda (to name a few). Framing AI as an independent, intelligent authority that just gives us final answers risks bypassing the increasingly critical aesthetic dimension, to say nothing of questioning the underlying assumptions built into each model. Perhaps, as we saw with writing, offsetting the negative dimensions of a new, powerful medium like AI will depend on developing novel forms of creativity, expression, and interpretation through the medium. This is important since we are going to inevitably lose some skills that were crucial with older mediums, but will not be critical for the new medium. Creating productive challenges within a new medium ensures students, for example, are still learning the critical thinking skills for work within that medium. After pocket calculators were invented, teachers focused more on proofs in the classroom and the development of mathematical thinking, rather than an exclusive focus on rote memorization. By focusing on the creative ways of working with a new technology, we begin to appreciate its complexity and the challenges that come with mastering that complexity.

The type of creativity that LLMs are enabling is close to what Peirce has called the action of musing: a reverie-like, unconstrained state of ``pure play'' and engagement with ideas guided by one's whim and mood, without having a specific objective in mind \citep{peirce1908neglected}. Even though on the surface level this might sound like lazy work, computer scientist Kenneth Stanely in his book titled \textit{The Myth of the Objective: Why Greatness Cannot be Planned}, provides historical anecdotes and computer simulations showing that truly novel discoveries and breakthroughs are only achieved when one is merely engaged in ``serendipitous discovery and playful creativity'' \citep{stanley2015greatness}. Discoveries are usually not achieved by rigid logic and setting inflexible objectives, but by iteratively musing and carving out aesthetic dissonances to find the next optimal stepping stone. This also highlights why the brain's default mode is ongoing spontaneous activity: that is nature's way of always remaining flexible enough and playing to unfold and welcome serendipitous breakthroughs.

Engaging with and consuming the products of this type of creativity is going to inevitably shape our conscious experience. Drawing from Dewey's quote that we also mentioned earlier, ``As the painter places pigment upon the canvas, or imagines it placed there, his ideas and feeling are also ordered'', we observe that LLMs allow us to expand our imagination and place novel pigments on a novel canvas. Through this process, our conscious experience is going to be uniquely reshaped by the novel forms we discover through this medium. As the instrument becomes more sophisticated and allows a higher capacity for expression, play, and creativity, the artist transforms with it.

\section{Conclusions}

New technology can channel our experience and connect us to something larger than ourselves, as part of an ongoing dance of folding and unfolding. As much as we've focused on the generativity of externalized media and serendipitous discovery ``out there'' in the world, we've also seen how this process reshapes the inner experience of creativity and aesthetics. Writing prior to the advent of the internet and even the personal computer, the French Jesuit and paleontologist Pierre Teilhard de Chardin spoke of a hereditary-like exteriorization of human knowledge. Tellingly, he called this process ``involution,'' evoking the folding-back upon oneself that can enable self-discovery and in Teilhard's language, ``enrichment'' \citep{steinhart2008teilhard}.

For Teilhard, the evolutionary development of symbols enabled humans to reflect back on themselves as individuals and as a collective, in effect enabling a distributed sphere of thought enveloping planetary human dynamics. As a result of this symbolic self-reflection, humans acquired the ability to realize their shared vulnerabilities and understand their most persistent tensions and longings. This shared vulnerability fosters ``an immediacy, an intimacy and a realism'' that leads to compassion and collaborative risk-taking

If we are only to focus on technology itself: the text, the brain, the cell, we may miss the inner heart of generativity and shared understanding these novel technologies enable. That we experience this technology as magic may have much to do with our animistic leanings, as Abram contends. It may also have much to do with our fundamental valuing of creativity, seeing our artifacts as potentially alive since we want our creative aliveness channeled through them. In a sense, what is fresh and original is what develops consciousness the most, as we never step into the same river twice. Or, in the motivated spontaneity of neural dynamics, it is the novel variations that are unfolded through conscious experience.

Our goal has been to shift the discourse on consciousness from a focus on metaphysics of substance to processes and dynamics. Of course, many others have focused on processes. Famously, Whitehead explained: the ultimate metaphysical principle is ``the creative advance into novelty'' \citep{whitehead1929process}. The argument we have advanced is on how this creative advance unfolds through the dynamic shaping of experience. By focusing on the geometry afforded by folding and unfolding, we can see how new shapes come into play -- and the centrality of play -- in this process. This is not an argument for a Platonic realm of pure, ideal forms, so much as an argument for naturalizing the restless generativity of life and consciousness. Novel shapes and forms that are vital, that matter, stick around to form the scaffolding upon which further innovations can be developed. That this pattern extends from DNA to brains and writing -- that is, from molecules to networks and to the seemingly immaterial realm of symbols and information -- may ultimately hold important truths regarding the fundamental nature and significance of our most treasured artifacts, and consciousness itself.

By realizing that meaning both results from, and is extended by, a process of playful unfolding, we are given a new purchase toward understanding the nature of conscious experience. Instead of focusing on how hard the problem of consciousness is, we focus on how soft its development is, malleable to the incessant folding and unfolding of life's dynamics. Artifacts such as neural synapses, the written word, and perhaps now, LLMs, reflect and reshape inner experience, creating a generative feedback loop that enables more differentiated thoughts and feelings.

In effect, consciousness has always been about the art of creativity. As such, metaphysics should be equally concerned with the process underlying this creativity and the serendipity that makes this possible. This points to a reconsideration of consciousness studies from a more interdisciplinary dialogue. That perhaps the study of ``scientific'' phenomena such as DNA, neural activity, LLMs, can be validly approached as nature's artistic adventures. In a sense, creative innovation and subsequent re-creations are just that, life's longing for recreation.

Do not artists, poets, and musicians have as much to offer us regarding a study of consciousness as the neuroscientist in the laboratory? Perhaps, when it comes to the study of consciousness, it's time for us all to have a bit more fun.

\section*{Acknowledgements}

This work was supported by a grant from the Department of Veterans Affairs (IK2CX002457 to MSJ).

\section*{Declaration of Interests}

The authors declare that they have no competing interests.

\bibliography{references}

\end{document}